\documentclass[12pt]{article}
\usepackage{amssymb, amsmath, amscd}
\usepackage{graphicx}
\usepackage[latin1]{inputenc}
\usepackage{epsfig}
\usepackage[affil-it]{authblk}

\usepackage{makecell}

\usepackage{subfigure} 

\usepackage[table]{xcolor}

\usepackage{multicol}

\numberwithin{equation}{section}

\newcommand\y{\cellcolor{green!10}}

\begin{document}

\bibliographystyle{amsalpha}

\title{Homology computation of {large point clouds} using quantum annealing}

\author{Raouf Dridi%
  \thanks{\texttt{raouf.dridi@1qbit.com}} \
  and  
  Hedayat Alghassi%
  \thanks{\texttt{hedayat.alghassi@1qbit.com}} }
\affil{1QB Information Technologies (1QBit)\\
       458 -- 550 Burrard Street\\
       Vancouver, British Columbia \, V6C 2B5, Canada}

\date{}
\maketitle

\begin{abstract}	
~~\\
Homology is a tool in topological data analysis which measures the shape of the data. In many cases, 
these measurements translate into new insights  which are not available by other means.  
To compute homology, we  rely on mathematical constructions which scale exponentially with the size of the data. 
Therefore, for large point clouds, the computation is infeasible using classical computers.
	In this paper,  we present a quantum annealing pipeline for computation of homology of large point clouds.   
The pipeline takes as input a  graph approximating the given point cloud. 
It uses quantum annealing to compute a clique covering of the {graph} and then uses this cover to construct  a Mayer-Vietoris complex.
The pipeline terminates by performing a simplified homology computation of the Mayer-Vietoris complex. 
We have introduced three different clique coverings and their quantum annealing formulation.
Our pipeline scales polynomially 
     in the size of the data, once the covering step is solved.
To prove correctness of our algorithm, 
we have also included tests using D-Wave 2X quantum processor. 

~\\


\end{abstract}

 
\newpage

\section{Introduction}
The abundance of data, of all sorts, represents  undoubtedly an exceptional and unprecedented 
wealth of knowledge for humanity to benefit from.  Yet, the extent of such abundance combined with the inherent complexity of the data make the deciphering and extraction of this knowledge tremendously difficult. Therefore  a data scientist faces two non
trivial challenges. First,  design models and algorithms appropriate to the complexity of the data and then leveraging them to large scales.  In our opinion, the appropriate algorithmics are to be found in advanced mathematics where concepts like ``correct glueing of local statistical information into a global insight''  are captured, precisely defined and solved.  Topological data analysis~(TDA) is one of these mathematics. It  uses algebraic topology; a branch of modern mathematics which investigates topological features through algebraic lenses. In this work,  
we leverage TDA algorithms to large scales using
quantum annealing.   

 ~~\\
The main concept in TDA is homology which 
is an invariant that consists of a sequence of vector spaces 
measuring the {shape} of a given point cloud.  These measurements usually  translate  into valuable insights about the data
which are not   available by other means. An excellent survey of TDA and its applications, in addition to 
a gentle introduction to the  notions of 
 homotopy equivalence, simplicial complexes and their homologies, 
  can be found in  \cite{zbMATH05545159} and \cite{Zomorodian:2012:AAC:2408007}; also we refer to  \cite{MR1095046} and \cite{zbMATH02103273} for more advanced reading.  The objective of the paper is to propose and test a quantum algorithm for computing the homology of large point clouds.
 
 ~~\\
Our approach rests on  {Mayer-Vietoris blow-up complex} \cite{Segal1968}. This approach starts, as any  homology computation method, with  approximating the given point cloud~$X$ with a  {simplicial complex}~$K$, as in Figure 1.   Commonly used complexes are
the so-called {witness } and Vietoris-Rips complexes, reviewed in the next section. 
    \begin{figure}[!htbp]\label{exples}
    \begin{minipage}{0.6\linewidth}
      \begin{center}
      \includegraphics[scale=0.2]{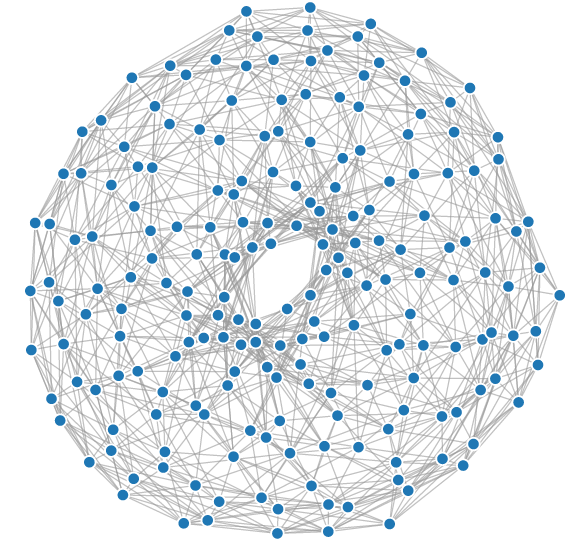}
      \end{center}
    \end{minipage}
    \hfill
    \begin{minipage}{0.4\linewidth}
      \includegraphics[scale=0.2]{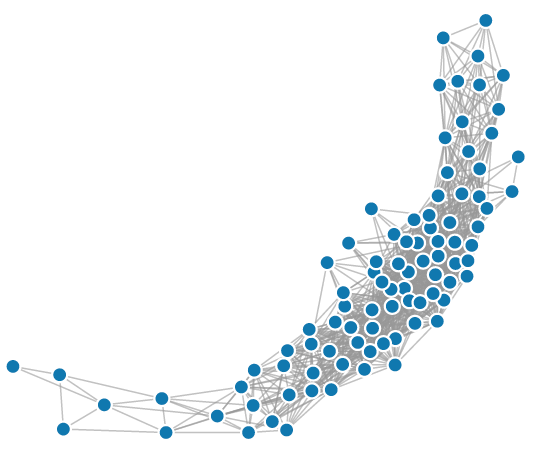}
    \end{minipage}
    \caption{{\small The left graph is the 1-skeleton of the witness complex of the unit torus with 100 landmarks and 100 000 data points. The corresponding witness complex is the 1-skeleton in addition to all its  cliques.   The right graph is the 1-skeleton  of the witness complex  (with 80 landmarks) of the NKI data set (\cite{doi:10.1056/NEJMoa021967}) that contains 24 496 gene expressions of 295 female patients diagnosed with breast cancer. } }
  \end{figure}
  ~~\\
Next, we cover the  
complex~$K$ with smaller subcomplexes {and then}  blow-up their overlaps. The blow-up operation 
 is a {homotopy equivalence} which implies that $K$ and it's {Mayer-Vietoris} complex have the same homology. 
Clearly, the efficiency of this construction  depends on the  covering step.
 
 ~~\\
Both Vietoris-Rips and witness complexes are clique complexes. Recall that a clique complex is an abstract complex given by the set of all cliques, of some graph (called 1-skeleton), sorted with the set theoretical inclusion.  In reality, we assume that only
 the 1-skeleton graph $G$ of $K$ is given  and 
 burden the homology pipeline with 
 the heavy task of constructing $K$, which is enumerating all cliques of $G$.

 ~~\\
The problem is now to cover  the 1-skeleton in a way which makes the subsequent computations of the pipeline (i.e., construction of $K$ and homology computation of the Mayer-Vietoris complex) scale ``nicely'' with the size of $G$. In the context of quantum computation, we seek  
exponential speed-up.   It is important to note that, in general, a brute force graph partitioning 
allows only  parallelization and computations still scale exponentially with size of the data.  

~~\\
We prove that the type of covering needed here is a clique based covering, that is, one needs to cover~$G$ with cliques.
The key points are: firstly,  cliques are homotopy equivalent to spheres, thus have knwon simple topology.  Secondly, a large portion of simplices 
of $K$ is  computed with this covering step i.e.,  a large fraction of simplices  are confined inside the covering cliques.  
 
 ~~\\
 We present three constructions for a such covering: 
minimum edge clique cover,  minimum vertex clique cover, and  an iterative method to compute a specific edge clique cover, that we call edge disjoint edge clique cover. 
We express the three methods as quadratic unconstrained binary optimization (QUBO) formulations \cite{QUBOhammer, Boros2002155}.  
%
We argue in the Discussion Section that our pipeline (simplicial complex construction and homology computation)  scales polynomially 
     in the size of the data, once the covering step is solved. The covering step  
     which is a binary optimization problem, can be solved via any optimization oracle
      thus, it is solver agnostic. Although we have tested our algorithm using 
      D-Wave 2X processor, any quantum annealing processor can be utilized.





 ~~\\
The present work is the first pipeline for homology computation using quantum annealing.   
A pipeline based on the gate model, has been recently proposed in \cite{seth}.  The key point there is
the compression of the simplicial complex $K$ into a quantum state in a $log_2(|K|)$-dimensional Hilbert space, spanned by the  simplices in $K$. Within this space, Betti numbers are computed  in polynomial time using quantum phase estimate. 
  It is interesting to mention that, in our paper, we also have some form of compression: the covering cliques  compress a lots of simplicial data. 

    ~~\\
 We usually track the  dimensions of the homology spaces (called Betti numbers) over a range of values of a persistent parameter $\epsilon$ (see definition next section). 
 We compute the so-called bar codes. Meaningful insights persist over a long range of  $\epsilon$; on the contrary, noise don't. On the other
 hand, a sudden change in the bar codes might point to an outlier. 
Real-life applications for a such pipeline would be 
subpopulation detection in cancer genomics, fraud detection in financial transactions  (\cite{ayasdi}),
brain networks  (\cite{5872535}), and robot sensor networks (\cite{SilvaGM05}), to name a few.   

     


 ~~\\
In Section 2.1, we review the witness and Vietoris-Rips complexes. In Section 2.2, we discuss Mayer-Vietoris complex and explain how the homology of this complex is computed. Section 2.3  contains our pipeline. The complexity analysis is discussed in Section 2.4.  We have included tests using the D-Wave 2X processor as well as a basic description of how such quantum processors work.

\section{Results}
\subsection{From data sets to simplicial complexes} \label{data2simp}
In order to compute homology algorithmically one needs to map the given data set into a simplicial complex. 
Here we give two conversion maps which are commonly used in TDA: the so called {\it Vietoris-Rips} and {\it witness complexes}. Before diving into their definitions, 
it is helpful
to give a bird's-eye view of the overall process. In general a mapping which 
assigns a simplicial complex to a data set is of the form 
\begin{eqnarray}\label{functor1}
	{\bf dataSets} &\rightarrow& {\bf simplicialComplexes}\\\nonumber
	X &\mapsto& K 
\end{eqnarray}
For the particular cases of  {Vietoris-Rips} and {witness complexes}, the map (\ref{functor1}) factors as 
\begin{eqnarray}\label{functor3}
	{\bf dataSets} &\rightarrow& {\bf Graphs} \rightarrow {\bf simplicialComplexes}\\\nonumber
	X &\mapsto & \quad G \quad   \quad \mapsto  K
\end{eqnarray}
where $K$ is the {clique complex} of $G$. 
 The computational  cost of this construction  is the highly non-trivial cost of enumerating all cliques in $G$.

~~\\
The difference between {Vietoris-Rips} and {witness complexes}  is in the definition of the graph $G$. 
The graph $G$, in the case of Vietoris-Rips complexes, is the neighbourhood graph: two points in the data set are connected if their distance is less than some parameter $\epsilon>0$.
In the case of witness complexes,  $G$ is defined as follows \cite{DeSilva:2004:TEU:2386332.2386359}.  Suppose we are given a set $\mathcal L\subset X$, called the landmark set.
 For every point $x\in X$, we let~$m_x$ denote the distance from this point to the set $\mathcal L$, i.e., $m_x =\mathrm{min }_{ l\in \mathcal L}\{d(x, l)\}$.  
{The graph $G$ is the graph whose vertex set is $\mathcal L$, and where a pair $\{ l, l'\}\subset \mathcal L$ is an edge if and only if there is a point $x\in X$ (the witness) such that ${d(x, \{ l, l'\})\leq m_x +\epsilon}$.  }  Finding the ``right" set 
 of landmarks is discussed in \cite{DeSilva:2004:TEU:2386332.2386359}. 
 
~~\\
Homology is   another mapping 
\begin{eqnarray}\label{functor2}
	 {\bf simplicialComplexes}&\rightarrow& {\bf AbelianGroups}\\\nonumber
	K &\mapsto& H_*(K)
\end{eqnarray}
which comes after (\ref{functor3}) and  assigns the Abelian groups $H_*(K) = \{H_0(K),\, H_1(K),\,  \cdots  \}$ to the simplicial complex $K$.
Assuming either witness, or Vietoris-Rips complexes are being, 
any homology computation pipeline must implement the two steps (\ref{functor3}) and (\ref{functor2}).   In order to be efficient, the pipeline must also scale 
``nicely'' with the size of the data set.

\subsection{Mayer-Vietoris blow-up complex}\label{parallel}
We recall here the definition of Mayer-Vietoris blow-up complex 
and then describe its homology computation. For the convenience of the non-expert reader, we have presented
this technical section through a simple example.

~~\\
  Let $K$ be a simplicial complex and suppose $\mathcal C= \{K^i\}_{i\in I}$ is a cover of  $K$ by simplicial subcomplexes  $K^i\subseteq K$, that is, $K=\cup_{i\in I}K_i$.  For $J\subseteq I,$ we define $K^J=\cap_{j\in J} K^j.$ 
The {\it Mayer-Vietoris blow-up complex} (\cite{Segal1968, Zomorodian2008126})  of the simplicial complex $K$ and cover $\mathcal C$ is defined by
\begin{equation}\label{MVcomplex}
	K^\mathcal C = \bigcup _{J\subseteq I} \bigcup_{ \sigma\in K^J}  \sigma\times J.
\end{equation}
A basis for the $k$-chains $C_k(K^\mathcal C)$ is {$\{ \sigma\otimes J \in K^\mathcal C |\, \mathrm{dim}\,  \sigma + \mathrm{card}\, J= n \}$}. The boundary of a cell $ \sigma\otimes J$ is given by:
$	
{	\partial ( \sigma\otimes J) = \partial  \sigma\otimes J + (\text{-}1)^{\mathrm{dim}\,  \sigma}  \sigma\otimes \partial J}.
$
We will not provide a proof here, but it is a fact that the projection $K^\mathcal C\rightarrow K$ is a homotopy equivalence and  induces 
an isomorphism $H_*(K^\mathcal C)\simeq H_*(K)$.

~~\\
The definition above boils down to the following: The simplicial complex $K^\mathcal C$ is  the set of ``original" simplices  in addition to the ones we get by blowing up common simplices.
These are of the form $\sigma\otimes J$ in the definition  above.  
In Figure~2, the vertex  $d$ common to the two subcomplexes $\{K_1, K_2\}$ is blown up into an edge $v\otimes 12$ and the edge $bc$ is blown up into the triangle-like
$bc\otimes 01$. In Figure 3, 
the vertex $a$ common to three subcomplexes $\{K_0, K_1, K_2\}$ is blown up into the triangle $a\otimes 012$. \\
\begin{figure}[!htbp]\label{bl1}
\begin{center}
\centerline{
\includegraphics[scale=0.4]{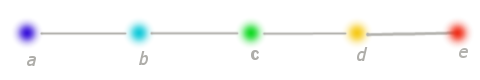}}
\centerline{
\includegraphics[scale=0.4]{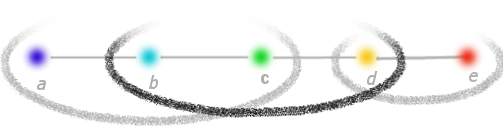}}
\centerline{
\includegraphics[scale=0.4]{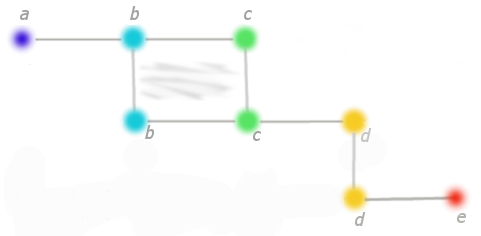}}
\caption{\emph{Top:} The simplicial complex $K$ is the depicted graph.
\emph{Middle:} $K$ is covered with $K_0, \, K_1$, and $K_2$.
\emph{Bottom:} The blow-up complex of the cover depicted in the middle image.  After the blow-up, the edges $b\otimes 01, \, c\otimes 01, \, d\otimes 12$, and the 2-simplex
$bc\otimes 01$ appear.}
\label{icml-historical}
\end{center}
\end{figure}  
\begin{figure}[!htbp]\label{bl2}
\begin{center}
\centerline{
\includegraphics[scale=0.35]{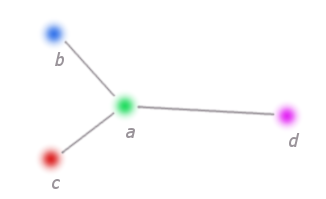}}
\centerline{
\includegraphics[scale=0.35]{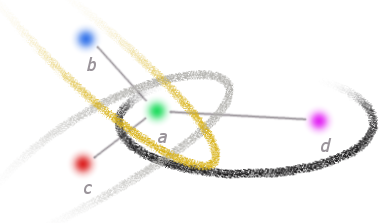}}
\centerline{
\includegraphics[scale=0.35]{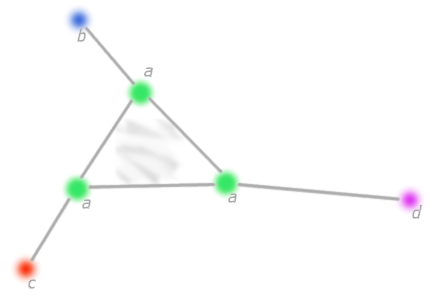}}
\caption{The triangle $a\otimes 012$ appears after blowing up the cover of the middle picture.}
\label{icml-historical}
\end{center}
\end{figure} \\
{Now, the key point  is that the boundary map of the simplicial complex $K^\mathcal C$ (which replaces $K$ by the homotopy equivalence) has a block form suitable for parallel rank
computation.}
As an example, let us consider   again 
the simplicial complex $K$ depicted in Figure 2. First, the space $C_0(K^\mathcal C)$ is spanned by the vertices 
$a\otimes 0,\, b\otimes 0,\, c\otimes 0,\, b\otimes1, \,c\otimes 1, \,d\otimes 1, \,d\otimes 2, \,e\otimes 2.$
That is, all vertices of $K$ take into account the partition to which they belong.
The space of edges $C_1(K^\mathcal C)$ is spanned by 
$ ab\otimes 0,  bc\otimes0, bc\otimes 1, cd\otimes 1, de\otimes 2,b\otimes 01,  \, c\otimes 01, \, d\otimes 12.$
That is, first the ``original" edges (i.e., those of the form $ \sigma\otimes j$, with $j\in J=\{0, 1, 2\}$ and $ \sigma$ being an edge in $K$) are constructed. Then, 
the new edges that result from blow-ups (i.e., those of the form $v\otimes ij$, where $v$ is a vertex
in $K^j\cap K^j$; if the intersection is empty the value of boundary map is 0), are constructed.
The matrix $\partial_0$ with respect to the given ordering is then:

 $$
\left(
\begin{array}{cc|cc|c|ccc}
\rowcolor{red!20}
  -1 & 0 & 0 & 0 & 0& 0&  0& 0 \\ 
  \rowcolor{red!20}
   1 & -1 & 0 & 0 & 0&-1& 0&0\\
   \rowcolor{red!20}
   0 & 1 & 0 & 0 & 0 & 0& -1&0\\
    \Xhline{3\arrayrulewidth}
    \rowcolor{green!20}
   0 & 0 & -1& 0 & 0& 1& 0& 0 \\ 
   \rowcolor{green!20}
  0 & 0 & 1 & -1 & 0&0& 1&0\\
  \rowcolor{green!20}
   0 & 0 & 0 & 1 & 0&0& 0&-1\\
    \Xhline{3\arrayrulewidth}
    \rowcolor{blue!20}
     0 & 0 & 0 & 0 & -1&0& 0&1\\
     \rowcolor{blue!20}
     0 & 0 & 0 & 0 & 1&0& 0&0\\
\end{array}
\right)
 $$
We see that one can now row-reduce each coloured block independently. There might be {\it remainders}, i.e., zero rows except for the intersection part. We collect all such rows in one extra matrix and row-reduce it at the end and aggregate. 
 For the second boundary matrix, we need to determine $C_2(K^\mathcal C)$. The 2-simplices
 are of three forms: the original ones (those of the form $ \sigma\otimes j$ with $ \sigma \in C_2(K)$; in this example there are none); those of the form $ \sigma\times \{i, j\}$, with $ \sigma \in K^i\cap K^j$; and those of the form $v\otimes \{i, j, k\}$, with $v\in K^i\cap K^j\cap K^k$ (there are none in this example but Figure 3 has one). We get 
 $
 {C_2(K^\mathcal C) =\langle bc\otimes 01\rangle}
 $. 


\subsection{The quantum pipeline}
This is the part of the paper where we describe our quantum pipeline which rests on Mayer-Vietoris construction we reviewed in the previous section.  We assume that we have been given a large data set $X$ and we have used one of the mapping procedures of Section \ref{data2simp} to map $X$ into a graph ${G}$. The (to be found) simplicial complex $K$, approximating $X$,
is the clique complex of ${G}$. 

~~\\
The key point in our pipeline is finding an {\it optimal} cover of $K,$
where  the subcomplexes $K_i$ are homotopy equivalent to a sphere\footnote{Similar to the covering step in \v{C}ech construction but the cover will be used in Mayer-Vietoris construction instead.}.
This is the same requirement as
covering the graph ${G}$ with cliques. It is  optimal because: 1) it reduces the rank computation, we have discussed in the previous section,  to only the intersection part, which is itself minimized, 2) it minimizes the construction of the simplicial subcomplexes $K_i$  (a consequence of the fact that the covering cliques confine a lots of simplicial data).    We formulate three different QUBOs for finding a such cover, all of them  can be implemented on any quantum annealing processor.

~~\\
The outline of this part is as follows.   In Section \ref{sub2}, we present our three different QUBOs, starting with minimum edge clique cover which is 
more complete than the other two in a sense that it covers all the edges.  This QUBO, however, needs many binary variables (binary variables for vertices and edges in each solution). Due to the practical limitation of qubit resources, we present two alternatives, minimum vertex clique cover (binary variables for vertices only in each solution) and  an iterative edge disjoint-edge clique cover method (only one set of binary variables for vertices).  The disjoint-edge clique cover does cover
all the edges but uses the quantum annealer more than once.    Later in Sections
\ref{simpConst} and \ref{localrk}, 
 we discuss the simplified simplicial complex construction from the given set of cliques and local rank computation for these three covering methods. Complexity analysis is detailed in the Discussion section.
 For completeness, we have also presented a QUBO formulation of the (minimum $k$-cut) graph partitioning (see  {\it Supplementary materials subsection 4.1}). 
 The subgraphs in graph partitioning covering do not have any specific structure, thus their local rank computation can not be simplified and the homolgoy computation is still exponential. It also assumes $K$ is given which is  problematic.   
We conclude in Section \ref{dwaveres} by testing our pipeline with the only available quantum annealing processor D-Wave 2X processor, as proof of concept and correctness. 


\subsubsection{Covering step} \label{sub2}
 We denote by ${\bf V}({G})$ and ${\bf E}({G})$ the vertex set and edge set of~{G}, respectively, and also define $n := |{\bf V}({G})|$ and $m :=|{\bf E}({G})|$. 

 ~~\\
 {\it 1. Edge clique cover.} 
The {first method} uses Edge Clique Cover (ECC), which is one of   Karp's 21 NP-complete problems (\cite{karp}). The problem is to cover
the set of edges of ${G}$ using a given number of cliques $k$. 
 For each (clique) solution $U_i$ , $ 0\leq i \leq k-1$, in the graph we need $n$ binary decision variables to represent vertices. The row vector ${\bf x}_i = (x_{i1}, x_{i2}, \dots, x_{in})$ is the solution vector that indicates if the vertex $v_j$ of $G$ belongs to the $i$th  clique solution $U_i$, the $x_{ij}$ is equal to 1; otherwise it is 0.   Similarly each row vector ${\bf e}_i = (e_{i1}, e_{i2}, \dots, e_{im})$ is the solution vector indicating if each of the $m$ edges are in the clique solution $U_i$. Let's define a $k(n+m)$ size binary variable vector $\mathbf{X}$:


\[{\mathbf{X}} = {\left[ {\begin{array}{*{20}{c}}
  {{{\mathbf{x}}_0}}&{{{\mathbf{x}}_1}}& \cdots &{{{\mathbf{x}}_{k-1}}}&{{{\mathbf{e}}_0}}&{{{\mathbf{e}}_1}}& \cdots &{{{\mathbf{e}}_{k-1}}} 
\end{array}} \right]^{\mathbf{T}}}\]
The edge clique cover QUBO formulation is then {\cite{hedayat}}:

\[\begin{gathered}
  \begin{array}{*{20}{c}}
  {\begin{array}{*{20}{c}}
  {}&{}&{}&{}&{}&{}&{}
\end{array}min}&{{{\mathbf{X}}^{\mathbf{T}}}{{\mathbf{Q}}_{{\mathbf{    }}}}{\mathbf{X}}} 
\end{array} \hfill \\
  {{\mathbf{Q}}_{{\mathbf{    }}}} = \left[ {\begin{array}{*{20}{c}}
  {{{\mathbf{I}}_k} \otimes ({{\mathbf{J}}_n} - {{\mathbf{I}}_n})}&{ - 2{{\mathbf{I}}_k} \otimes {\mathbf{B}}} \\ 
  { - 2{{\mathbf{I}}_k} \otimes {{\mathbf{B}}^T}}&{\left( {{{\mathbf{J}}_k} + 3{{\mathbf{I}}_k}} \right) \otimes {{\mathbf{I}}_m}} 
\end{array}} \right] \hfill \\ 
\end{gathered} \]
where  $\mathbf{B}$ is the incidence matrix of   $G$. 
The matrix  
$\mathbf I_k$ is the {$k\times k$} identity matrix and $\mathbf J_k$ is  the {$k\times k$} one matrix with all entries 1.
 Here, for simplicity, we have bounded the number of cliques that can cover an edge to two.
Thus $k(n+m)$ binary variables is required. To increase the bound on number of overlapping cliques (that can cover an edge), we need to add extra slack variables as many as $\left\lceil {\frac{k}{2}} \right\rceil m$.  
 The smallest number of cliques that cover ${\bf E}({G})$ is called edge clique cover number or intersection number $\theta_e(G)$. An upper bound for $\theta_e(G)$ is  $2e^2(d+1)^2 log_e(n)$ with $d = \Delta (\overline {G} )$ is the maximum degree of complement of {G} \cite{Alon}. If we assume the graph is dense, which is the case for many hard homology problems,  $d$ would be small and the intersection number is on the order of~{$log(n)$.}

 ~~\\
{\it 2. Vertex clique cover.}  An approach with lower number of variables, is to compute a Vertex Clique Cover (VCC) for the graph $G$. This problem, which is also among the Karp's 21 NP-complete problems \cite{karp},  consists of covering the vertex set with cliques such that each vertex is assigned to a unique clique. It can be translated into a vertex colouring problem where we cover $\overline {G}$ with a given number of colours such that no edge connects two vertices of the same colour.     Let ${\mathbf{A}}$ be the adjacency matrix of the graph ${G}$, and $\mathbf{X}$  denotes:
\[{\mathbf{X}} = {\left[ {\begin{array}{*{20}{c}}
  {{{\mathbf{x}}_0}}&{{{\mathbf{x}}_1}}& \cdots &{{{\mathbf{x}}_{k-1}}} 
\end{array}} \right]^{\mathbf{T}}}\]
the QUBO problem formulation of the vertex clique cover is {(\cite{hedayat})}:

\[\begin{array}{*{20}{c}}
  {\begin{array}{*{20}{c}}
  {min}&{{{\mathbf{X}}^{\mathbf{T}}}\left( {{{\mathbf{Q}}_{{\mathbf{main}}}} + \alpha {{\mathbf{Q}}_{{\mathbf{orth}}}}} \right){\mathbf{X}}} 
\end{array}} \\ 
  \begin{gathered}
  {{\mathbf{Q}}_{{\mathbf{main}}}} = {{\mathbf{I}}_k} \otimes \left( {{{\mathbf{J}}_n} - {{\mathbf{I}}_n} - {\mathbf{A}}} \right) \hfill \\
  {{\mathbf{Q}}_{{\mathbf{orth}}}} = ({{\mathbf{J}}_k} - 2{{\mathbf{I}}_k}) \otimes {{\mathbf{I}}_n} \hfill \\ 
\end{gathered}  
\end{array}\]
where $k$, the number of cliques in the problem, is chosen greater than or equal to the clique covering number of  $G$, $\theta({G})$ (equal to $\chi(\overline {G})$, the chromatic number of $\overline {G}$).   An upper bound for $\theta(G)$ is $d = \delta ({G})$, which is the minimum degree of graph $G$ (Brooks' theorem \cite{Brooks}).

~~\\
{\it 3. Edge disjoint-edge clique cover. } 
The third clique covering method is a variation of the edge clique cover,  we call it Edge Disjoint-Edge Clique Cover (ED-ECC). Here, the covering subgraphs  intersect only at vertices.   
The algorithm takes as input the graph~${G}$ and a stopping criterion. The idea is to iteratively find the maximum clique and each time remove the clique edges from the graph of the previous iteration.  Each run gives one maximum clique. At step~{  $i$}, we get a new graph {   ${G}_i$} 
with adjacency matrix  ${\bf A}_i$.  We stop when the clique is  small (stopping {criterion 1}) or after a certain number of cliques computation (stopping {criterion~2}). 
The QUBO formulation for finding the maximum clique (at iteration $i$) is  
$$
	min\quad  {\bf x}^T  ({\bf A}^{(i)}-{\bf I}_n) {\bf x},
$$
where {\bf x} $= (x_1, \dots, x_n)^T,$ 
 ${\bf A}^{(i)}$ is the updated adjacency matrix at step $i$, and $n$ is the dimension of~${\bf A}^{(i)}$. The adjacency matrix of the maximum
 clique is then $C^{(i)} = {\bf x}^{(i)} {{\bf x}^{(i)}}^T  \circ ({\bf J}_n -{\bf I}_n)$, where $\circ$ is the Hadamard product and  ${\bf x}^{(i)}$ is the solution of the QUBO problem at iteration $i$.   There is an obvious gain in terms of the size of the problems that we can handle. Indeed, the number of variables involved here is only $n$, making this covering method more practical considering current limitation on the size of quantum annealing processor.

\subsubsection{Construction of the Mayer-Vietoris complex}\label{simpConst}
We describe now how the Mayer-Vietoris complex is constructed for the three different covers we presented above.  
A large number of the simplices are confined inside the covering cliques. One needs, however, to find the few remaining simplices
outside these covering cliques, depending on the covering method. 
  The set of these remaining simplices will be the last subcomplex $K_k$ in the cover
   $K=\cup_{i\in { \{0, \cdots, k\}}}\left\{K_i \right\}$. 
   The subcomplexes $K_i$, for $0\leq i\leq k-1$, are the power set of the clique $U_i$. 

~~\\
For the edge clique cover (and Edge disjoint-edge clique cover),  the last subcomplex  $K_k$ is the clique complex of  the subgraph $U_k$ defined as follows. Its vertex set ${\bf V}(U_k)$ is the set of vertices inside the pair-wise intersections between the covering cliques $\{U_i \}_{i=0...k-1}$. The edge set ${\bf E}(U_k)$ is  the restriction of ${\bf E}({G})$ to ${\bf V}(U_k)$.   For the vertex clique cover,  the last subcomplex $K_k$ is the clique complex of the graph $U_k$ whose vertex set ${\bf V}(U_k)$ is the set of all the vertices of the connecting edges and its set of edges ${\bf E}(U_k)$ is the restriction of ${\bf E}({G})$ to  ${\bf V}(U_k)$.  

~~\\
The complexity of
this step is dictated by the size of the intersections; since one needs to connect the vertices in these intersections. This is 
the same as analyzing the size of the matrices $\mathcal B_i$ we introduce next.  

\subsubsection{The rank calculation}\label{localrk}
As described in  Section \ref{parallel} 
the boundary matrix of the Mayer-Vietoris complex,  independently of the nature of the cover we choose,  has the form
\begin{equation}\label{parallelmat}
\left(
\begin{array}{c|ccc|c}
\rowcolor{red!20}
  \y \mathcal A_0&  0& \cdots &0& \y \mathcal B_0 \\ 
    \Xhline{3\arrayrulewidth}
    \rowcolor{red!20}
     0 &\y \mathcal A_1&   \cdots &0& \y \mathcal B_1 \\ 
    \Xhline{3\arrayrulewidth}
        \rowcolor{red!20}
     \vdots & \vdots&   \cdots &\vdots& \y \vdots \\ 
    \Xhline{3\arrayrulewidth}    
        \rowcolor{red!20}
     0 &0&   \cdots & \y \mathcal A_{k} & \y \mathcal B_{k} \\ 
    \Xhline{3\arrayrulewidth}
\end{array}
\right)
\end{equation}
where the matrix  $\mathcal A_i$ is the boundary matrix of the subcomplex $K_i$. 
In general, this will not yield any  speed-up and it only allows parallelization of the computation.  The situation changes substantially  using one of the clique based covers above. The fact that each of the $K_i$ for $i\in \{0, \cdots, k-1\}$ is   
homotopy equivalent to a sphere,  results in  
a reduced rank computation which scales polynomially with the size of the graph ${G}$
 (See Discussion). For $i\in \{0, \cdots, k-1\}$,   $\mathrm{rank}\, \mathcal A_i=\mathrm{rank}\, \partial_\ell^i= \sum_{\alpha=0...\ell}  (\mbox{-}1)^{\ell-\alpha}\binom  {m} {\alpha}$ with $m$ being the size of the clique $U_i$. Additionally, the passage matrix $\mathcal P_i$, which makes $\mathcal A_i$
upper triangular ($\mathcal A_i\mathcal P_i$ is upper triangular) is also known. 
To find the remainders (see Section \ref{parallel}) we let 
 $r_i =\mathrm{rank}\, \mathcal A_i$
be the precomputed rank of~$\mathcal A_i$. The remainder is then given by the product $\mathcal B_i[r_i + 1, end] \mathcal P_i$,
where~${\mathcal B_i[r_i+1, end]}$ is the submatrix of $\mathcal B_i$ containing the rows~$r_i + 1$ downward.  

~~\\
In the case of disjoint edge clique cover, 
by construction, the covering subcomplexes~${\{K_i\}_{i=0...k}}$  intersect only at vertices (however, a vertex can be blown up
into a high-dimensional simplex if it belongs to several covering subgraphs). This translates into a considerable  reduction in the size of the 
matrices~$\mathcal B_i$ which now involve only simplices of the form $v\otimes J$ , where~$v$ is a vertex in ${G}$ and $J$
is the set of all subcomplexes containing the vertex~$v$. 
Obviously, the complexity of the computation is defined by the dimension of the submatrices~$\mathcal B_i$ (See Discussion Section).


%


\subsubsection{Implementation on D-Wave quantum processor}\label{dwaveres}
We have tested our algorithm on the D-Wave 2X machine over many instances of the  solid torus, as a proof of concept. Here, we report
some statistics. 
Due to the embedding  limitations  of the D-Wave 2X processor (Appendix A.2), some of the instances were not successfully embedded into the processor, thus we could not calculate their Betti numbers.
Columns of tables below represent 1) $n$ the number of vertices, 2) $m$ the number of edges, 3) $D$ the density of the graph ,  4) the intersection number:   $\theta _e(G)$ in ECC and  $\theta(G)$ in VCC,  5) $|P|$ the problem size  (i.e., the number of binary variables in the QUBO), 6)  $Embed$ the embedding and solving status   and  7) $Betti$ the Betti number calculated status. The samples are sorted based on number of vertices and problem size, so the reader can see the border of embeddable graphs for each method. Since some of the QUBO's are more sparse than others, they can be embedded in higher problem sizes.

\begin{center}
\[\begin{gathered}
  \mathop {\boxed{\begin{array}{*{20}{c}}
n&m&D&{{\theta _e}}&{\left| P \right|}&{Embed}&{Betti} \\ 
6&{13}&{0.87}&4&{76}&\surd &\surd  \\ 
8&{12}&{0.43}&4&{80}&\surd &\surd  \\ 
8&{20}&{0.71}&4&{112}& \times &{NA} \\ 
{10}&{15}&{0.33}&5&{125}& \times &{NA} \\ 
{12}&{24}&{0.30}&4&{144}& \times &{NA} 
\end{array}}}\limits^{{\mathbf{ECC}}} 
\mathop {\boxed{\begin{array}{*{20}{c}}
n&m&D&\theta &{\left| P \right|}&{Embed}&{Betti} \\ 
{12}&{36}&{0.55}&3&{36}&\surd &\surd  \\ 
{12}&{24}&{0.36}&4&{48}&\surd &\surd  \\ 
{15}&{45}&{0.43}&4&{60}&\surd &\surd  \\ 
{16}&{40}&{0.33}&4&{64}& \times &{NA} \\ 
{16}&{56}&{0.47}&4&{64}& \times &{NA} 
\end{array}}}\limits^{{\mathbf{VCC}}}
\end{gathered}
\]
\[\begin{gathered}
\mathop {\boxed{\begin{array}{*{20}{c}}
n&m&D&{\left| P \right|}&{Embed}&{Betti} \\ 
{40}&{580}&{0.75}&{40}&\surd &\surd  \\ 
{50}&{725}&{0.59}&{50}&\surd &\surd  \\ 
{60}&{1320}&{0.75}&{60}&\surd &\surd  \\ 
{70}&{1435}&{0.59}&{70}& \times &{NA} \\ 
{80}&{1560}&{0.47}&{80}& \times &{NA} 
\end{array}}}\limits^{{\mathbf{ED - ECC}}} 
\end{gathered} \]
\end{center}%
~~\\
For all instances that the problem was successfully embedded into D-Wave 2X processor, our algorithm successfully calculated the Betti numbers. 
Note that we only observed the minimum energy solution, among many reads of each annealing process, since our task is to prove the concept.  The reader should also note that, these tests only show the correctness of each method's implemented algorithm. The performance comparison and scaling characteristics of discussed algorithms cannot be evaluated with current size of quantum annealing processor.

\section{Discussion}\label{discussion}
In this paper we have discussed how quantum annealing can be used to speed-up the homology computation of large point clouds
and presented proof-of-concept tests using the newest D-Wave 2X quantum processor. 
Additionally, we have presented our work as a complete data mining pipeline.

 ~~\\
Our pipeline is dedicated to dense  graphs; sparse cases  should be treated classically.  Clearly, the complexity of the pipeline is defined by the dimension of the submatrices~$\mathcal B_i$ of the boundary map matrix (\ref{parallelmat}). 
This dimension  is given by the number of the blownup simplices, that is, simplices of the form $\sigma\otimes J$ with $|J|>1$. Precisely, to compute the $l^{th}$ Betti number $\beta_\ell$, we count all $\ell+1$ simplices of the form  $\sigma\otimes J$, such that $dim(\sigma) + |J|-1 =\ell+1$ in addition to $|J|>1$.  
For the vertex and edge cover, the intersection  between the subgraphs $\left\{U_j \right\}_{j=0..k}$  is always a clique. 
The column dimension of $\mathcal B_i$, needed for $\beta_\ell$, would be then
$\sum_{i=0..\ell}  \dbinom{\omega}{\ell +1-i}  \dbinom{\kappa}{i+2}$
with~$\omega := \mathrm{max}\, \{|\cap_j U_j| \}$ the size of the maximum intersection and $\kappa$ 
is the maximum number of subgraph $U_j$ with non empty overlap (i.e., the size of the maximum simplex in the nerve complex $\mathcal N \{U_j\}$). 
Obviously,  $\kappa$  is  less than $\theta_e(G)$ which, for dense graph, is in the order of $O(log(n))$ in the edge clique cover case. 
If $\kappa$ is very big then the subgraphs $U_j$ are small and thus $\ell$ is small 
($\beta_\ell$ is also the Betti number of 
the initial witness complex $K$ and thus $\ell$ is limited by the size of the maximum subgraph). 
If $\omega$ is very big then the graph is covered with only a small number of cliques. Almost all  other cliques
are confined inside these covering cliques and thus in the image of the boundary map. 
This implies that, after we mod out by the image of the boundary map,  we dont have enough simplices to bound high dimensional voids and thus $\ell$ is small.
The conclusion of this is that our algorithm, using edge clique cover,  is polynomial in time 
for graphs in which $\kappa$ and $\omega$ are not very big. These are the type of graphs which are intractable 
classically. 
For vertex clique, the argument and conclusion are the same with replacing $\theta_e(G)$ with $\theta(G)$.  By Brooks' theorem, $\theta(G)$ is bounded by the degree of the complement graph. And since the complement is sparse, $\theta(G)$ is   small. 
For edge disjoint edge clique covering,  the column dimension of~$\mathcal B_i$ reduces to 
$\omega\times \nu $  with $\nu := \mathrm{card} \{J\subset \{0, \cdots, k\}\, |\, |J|=\ell+2,\,  \cap_{j\in J} C_j \neq \phi\}$, since the intersection is forced to be at the vertex level only. In this case, our algorithm  takes $O(\omega^2\times \nu^2$) i.e., polynomial in the size of the graph since $\omega <n$. 



\newpage

\section{Supplementary materials}

\subsection{Partitioning based parallelization of the homology computation}  
Partitioning the graph $G$ using the minimum $k$-cut will not yield any significant speed-up and the computation is still exponentially expensive, it only allows parallelization.  Additionally, it assumes
the simplicial complex $K$ given which is problematic. We present it here for completeness (since it has been discussed in \cite{ryan1}, where METIS library \cite{Karypis:1998:FHQ:305219.305248} is used). 

~~\\
The  minimum $k$-cut is another NP-problem  in Karp's 21 list (\cite{karp}). It consists of 
partitioning the vertex set of $G$ into $k$ non-empty and fixed-sized subsets so that the total weight of edges connecting distinct subsets is minimized. The QUBO formulation of the minimum $k$-cut problem is formulated in (\cite{fanneng})  (see also\cite{hedayat}).
 For each partition $U_i$ in the graph, for $ 0\leq i \leq k-1$, we need $k$ binary decision vectors of size $n$. The row vector ${\bf x}_i = (x_{i1}, x_{i2}, \dots, x_{in})$ is part of the solution that indicates that if the vertex $v_j$ of $G$ belongs to the $i$th  partition, the $x_{ij}$ is equal to 1; otherwise it is 0.   Let  $\mathbf{X}$  denotes:
\[{\mathbf{X}} = {\left[ {\begin{array}{*{20}{c}}
  {{{\mathbf{x}}_0}}&{{{\mathbf{x}}_1}}& \cdots &{{{\mathbf{x}}_{k-1}}} 
\end{array}} \right]^{\mathbf{T}}}\]
The minimum $k$-cut graph partitioning problem is  then the QUBO:
\[\begin{array}{*{20}{c}}
  {\begin{array}{*{20}{c}}
  {min}&{{{\mathbf{X}}^{\mathbf{T}}}\left( {{{\mathbf{Q}}_{{\mathbf{main}}}} + \alpha {{\mathbf{Q}}_{{\mathbf{orth}}}} + \beta {{\mathbf{Q}}_{{\mathbf{card}}}}} \right){\mathbf{X}}} 
\end{array}} \\ 
  \begin{gathered}
  {{\mathbf{Q}}_{{\mathbf{main}}}} =  - {{\mathbf{I}}_k} \otimes {\mathbf{A}} \hfill \\
  {{\mathbf{Q}}_{{\mathbf{orth}}}} = ({{\mathbf{J}}_k} - 2{{\mathbf{I}}_k}) \otimes {{\mathbf{I}}_n} \hfill \\
  {{\mathbf{Q}}_{{\mathbf{card}}}} = {{\mathbf{I}}_k} \otimes \left( {{{\mathbf{J}}_n} - 2{s_{av}}{{\mathbf{I}}_n}} \right) \hfill \\ 
\end{gathered}  
\end{array}\]
The matrix  
$\mathbf I_k$ is the {$k\times k$} identity matrix, $\mathbf J_k$ is  the {$k\times k$} one matrix with all entries 1, 
${s_{av}} = \frac{1}{2}\left( {\left\lceil {\frac{n}{k}} \right\rceil  + \left\lfloor {\frac{n}{k}} \right\rfloor } \right)$
represents the average size (cardinality) of partitions.  Also $\alpha$ and $\beta$ are the orthogonality and cardinality constraint balancing factors.  

~\\
Once we pass this QUBO to the quantum annealer we obtain subgraphs $\{U_i\}_{0\leq i \leq k-1}$ which, in addition to an extra subgraph containing all of the edges between them (now minimized), define a covering for $G$.
To execute the parallel computation, we need to complete this graph covering into a cover 
of $K$ in terms of subcomplexes for which we have~${K=\cup_{0}^{k}\, K_i}$. For this, we assign the subcomplex $K_i$  to the simplex $\sigma\in K$  if its vertices belong to the subgraph~$U_i$. Otherwise, the simplex
is put in the extra cover~$K_{k}$. Similar to the tables in Section \ref{dwaveres}, the table below  shows some statistics
for the graph partitioning. 


\begin{center}
\[\begin{gathered}
 {\boxed{\begin{array}{*{20}{c}}
  n&m&D&k&{\left| P \right|}&{Embed}&{Betti} \\ 
{12}&{24}&{0.36}&4&{48}&\surd &\surd  \\ 
 {12}&{36}&{0.55}&4&{48}&\surd &\surd  \\ 
{12}&{48}&{0.73}&4&{48}&\surd &\surd  \\ 
{16}&{40}&{0.33}&4&{64}& \times &{NA} \\ 
{16}&{56}&{0.47}&4&{64}& \times &{NA} 
\end{array}}}
\end{gathered} \]
\end{center}

\subsection{Quantum annealing}
Here we introduce the quantum annealing concept that ultimately solves a general Ising (QUBO) problem, then  talk about the important topic of embedding a QUBO problem into the specific quantum annealer
(D-Wave 2X processor).

~~\\
 Quantum annealing (QA), along with the D-Wave processor, have been the focus of much research. We refer the interested reader to 
\cite{natueDwave, Calude:2015:GCA:2744447.2744459, BoixoNature, BoixoNature2, PhysRevX.4.021041}.  
QA is a paradigm designed to find the ground state of systems of interacting spins represented by a time-evolving Hamiltonian:
$$
	\mathcal S(s) =\mathcal E(s)\mathcal H_P - \frac{1}{2} \sum_i \Delta(s)\sigma_i^x,
$$
$$
	\mathcal H_P = -\sum_i h_i \sigma_i^x + \sum_{i<j}J_{ij}\sigma_i^z\sigma_j^z.
$$
The parameters $h_i$ and $J_{ij}$ encode the particular QUBO problem $P$ into its {Ising} formulation. QA is performed by first setting $\Delta \gg \mathcal E,$ which results in a ground state into which the spins can be easily initialized. Then $\Delta$ is {\it slowly} reduced and $\mathcal E$ is increased until $\mathcal E\gg \Delta$. At this point the system is dominated by $\mathcal H_P,$ which encodes the optimization problem. Thus the ground state represents the solution to the optimization problem.

\subsection{Embedding} 
An embedding is the mapping of the nodes of an input graph to the nodes of the destination graph. The graph representing the problem's QUBO matrix needs to be embedded into the actual physical qubits on the processor in order for it to solve the QUBO problem. The specific existing connectivity pattern of qubits in the D-Wave chip is called the Chimera graph. Embedding an input graph (a QUBO problem graph) into the hardware graph (the Chimera graph) 
is in general NP-hard~{(\cite{Choi}).} 

 \begin{figure}[!htbp]
\begin{center}
\begin{subfigure}
{\includegraphics[scale=0.45]{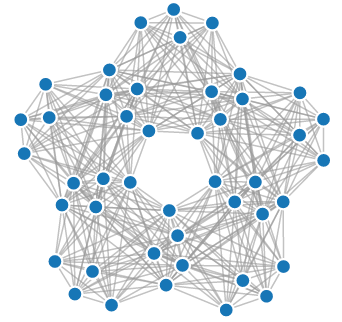}}
\end{subfigure}
\end{center}
\begin{center}
\begin{subfigure}
{\includegraphics[scale=0.45]{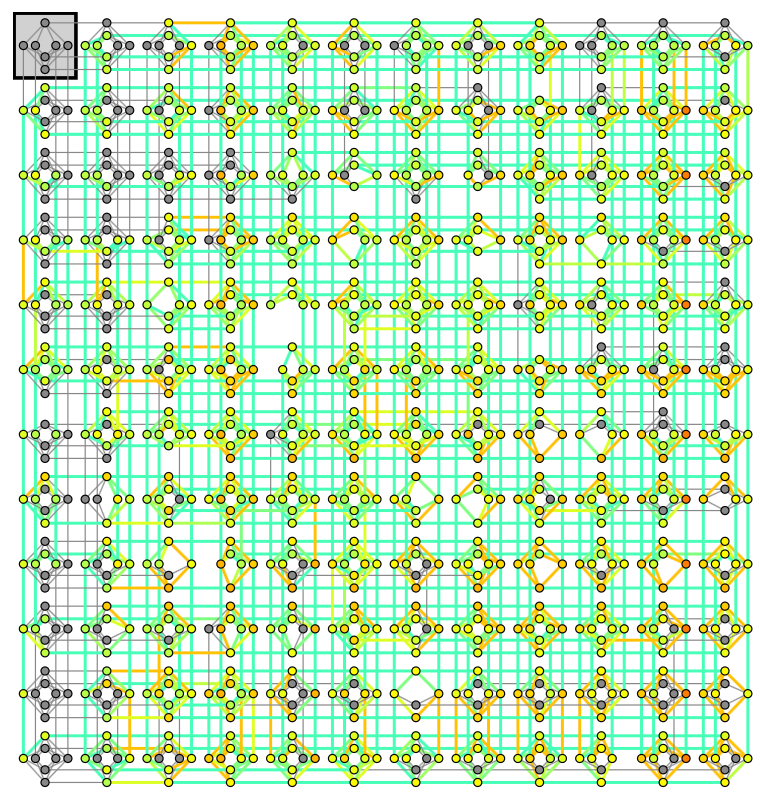}} 
\caption{\emph{Top:} A sample 1-skeleton, consists of five cliques of size 14 overlapping on five nodes.
\emph{Bottom:} The actual embedding of the corresponding edge disjoint edge clique cover problem inside the current D-Wave 2X Chimera graph.
The QUBO problem has 50 variables, 280 quadratic terms used to map the problem and 586 quadratic terms used to map chains.  The colouring  of the nodes (and edges, respectively) represents the $h$ parameter values (the $J$ values, respectively) according to the colouring scheme on the interval [-1, 1] represented by a colour interval [blue, red] in the D-Wave 2X processor API.}
\label{chimera}
\end{subfigure}
\end{center}
\end{figure}

 ~~\\
Figure \ref{chimera} shows a sample embedding into the Chimera graph of the D-Wave 2X chip consisting of an {${12 \times 12}$} lattice of {${4 \times 4}$} bipartite blocks. The Chimera graph is structured so that the vertical and horizontal couplers in its lattice are connected only to either side of each bipartite block.  Each node in this graph represents one qubit and each edge represents a coupling between two qubits. 
Adjacent nodes in the Chimera graph can be grouped together to form new effective (i.e., logical) nodes, creating nodes of a higher degree. Such a grouping is performed on the processor by setting the coupler between two qubits to a large negative value, forcing two Ising spins to align such that the two qubits end up with the same values. These effective qubits are expected to behave identically and remain in the same binary state at the time of measurement. The act of grouping adjacent qubits (hence forming new effective qubits) is called chain creation or identification. 
 
  ~~\\
  An embedding strategy consists of two tasks: mapping and identification.  Mapping is the assignment of the nodes of the input graph to the single or effective nodes of the destination graph. Solving such problems optimally is in general NP-hard, but one can devise various approximations and enhancement strategies to overcome these difficulties, for example, using statistical search methods like simulated annealing, structure-based methods, or a combination of both. For a better understanding of current embedding approaches, we refer the reader to \cite{Choi}, \cite{roy}, \cite{king0}, \cite{king1}. In Figure \ref{chimera} (bottom), the blue lines indicate the identified couplers, the yellow lines indicates the problem couplers (i.e., the edges of the problem graph), and the grey lines indicate empty couplers.
 
 \section*{Acknowledgements}
 {We would like to thank M. Elsheikh for constructive discussions about the parallelization, S. S. Rezaei  for his helpful comments,  P. Haghnegahdar, J. Oberoi and  P. Ronagh for helpful discussions on the topic, and M. Bucyk for proof reading the manuscript.}
 
\bibliography{c}
 
\end{document}